\documentclass[11pt]{article}

\usepackage{graphicx}
\begin{document}
\begin{center}
Chemical telemetry of OH observed to measure interstellar magnetic fields
\vskip 1cm

Serena Viti$^{1}$ Thomas W. Hartquist$^{2}$ and Philip C. Myers$^{3}$
\end{center}

\noindent
Department of Physics and Astronomy, University College London,
London WC1E 6BT, UK  {sv@star.ucl.ac.uk} \\
School of Physics and Astronomy, University of Leeds, Leeds LS2 9JT, UK  \\
Harvard - Smithsonian Center for Astrophysics, 60 Garden Street,
Cambridge MA 02138, USA \\
\date{Received; accepted }

\abstract{
We present models for the chemistry in gas moving towards the
ionization front of an HII region. When it is far from
the ionization front, the gas is highly depleted of elements more
massive than helium. However, as it approaches the ionization front,
ices are destroyed and species formed on the grain surfaces are
injected into the gas phase. Photodissociation removes gas phase
molecular species as the gas flows towards the ionization front. We
identify models for which the OH column densities are comparable to
those measured in observations undertaken to study the magnetic fields
in star forming regions and give results for the column densities of
other species that should be abundant if the observed OH arises
through a combination of the liberation of H$_2$O from surfaces and
photodissociation. They include CH$_3$OH, H$_2$CO, and H$_2$S. 
Observations of these other species may help
establish the nature of the OH spatial distribution in the clouds,
which is important for the interpretation of the magnetic field
results. \\

keywords ISM: HII regions, magnetic fields, molecules - masers - stars: formation

}

%

\section{Introduction}

   It is not yet clear whether clouds are generally magnetically
supercritical or subcritical in the sense that the mass-to-magnetic flux ratio
is above or below the critical value at which gravity overcomes the support
by the magnetic field (Mouschovias \& Spitzer 1976). What is clear, however, is
that this distinction is very important, as the mode of star formation
depends on whether a cloud is magnetically supercritical or subcritical
(e. g. Shu et al. 1987).  Measurements of interstellar magnetic field
strengths in molecular clouds are vital.

Crutcher (1999) produced a detailed compilation of reliable results,
obtained up to the time of his writing, for line - of - sight magnetic
field strengths and upper limits inferred from observations
of OH in a sample of molecular clouds. He also collected data on
column densities, line widths, number densities, temperatures, and
molecular cloud sizes. Crutcher (1999) concluded that the values of
the mass - to - magnetic flux ratios are about twice the critical
value. Shu et al. (1999) compared the data with a
model of a highly flattened cloud (Allen \& Shu 2000) and inferred
that the mass - to - flux ratios have values that are approximately
equal to the critical value.  Bourke et al. (2001) have obtained
further OH and column density data and performed a related analysis
and they concluded that their results are model dependent: if the model
cloud is initially a uniform sphere with a uniform field strength then
the cloud is magnetically supercritical. If instead, the model cloud
is a flattened sheet, the data imply that the system is 
stable.  In general, their detection rate was low and they explained
this as a result of a selection effect: most of their sources were
towards H II regions that are probably expanding; the expansion of an H II
region leads to the compression of the surrounding gas into a thin
shell-like structure, decreasing the Zeeman effect. This tends to make
clouds that are not necessarily magnetically supercritical appear as
though they are.

\par   The analysis to determine the dynamical significance of the measured
line - of - sight magnetic field strengths are based on the assumption
that the OH Zeeman data provide a measure of the magnetic field
strength throughout the region where most of the material,
contributing to the column density, is.  However, the possibility that
OH is particularly abundant in restricted shells around HII regions
has been suggested (Elitzur \& de Jong 1978; Hartquist et al. 1995)
in work on maser sources. Due to the spatial coincidence of OH and
CH$_3$OH maser sources in W3(OH) established by Menten et al. (1992),
Hartquist et al. (1995) argued that the release of icy mantles from
grain surfaces in gas moving towards an ionization front plays a major
role in the chemistry in the maser regions. CH$_3$OH is a direct
product of the surface chemistry, and OH is generated through the
photodissociation of H$_2$O injected into the gas phase from the
surfaces. Crutcher (2004, private communication) has argued that the absence of
limb brightening and other observational results demonstrate that in some
sources, for which the OH data yield clear measurements of the magnetic
field strength, the OH is not primarily concentrated in shells. However,
he does accept the possibility that in at least some of the sources
for which the studies yield only upper limits for the field strengths,
OH may be mostly in shells. This might contribute to the null detections
of the Zeeman effect in some sources.

The main motivation of this paper is an attempt to develop a means to establish whether in some cases OH, used in measurements of magnetic field strengths, exists in rather localized regions as a consequence of forming in photodissociation regions. 
Our study shows that it is indeed possible to produce OH with a 
column density
in the range of those measured by Bourke et al. (2001) in
the manner described by Hartquist et al. (1995). We have developed the
model in the first step in an attempt to chemically determine whether the
molecular gas around H II regions may indeed be swept up into thin shells,
resulting in a nonuniform magnetic field geometry, that leads to measured
values of the field strength that are much lower than the true values. If the
OH is formed as gas approaches the ionization front, our model allows
the prediction of the column densities of other species that may be
observed to confirm the validity of the model. Measured column densities
for these other species that disagree with the predictions would suggest
that the OH is not primarily in gas near the ionization front.

In section 2 we describe the model assumptions. Section 3 contains our
results and discussion, and Section 4 concludes the paper.

\section{Description of the Model}

The basic chemical model we adopt is a modification of the model
employed in Viti et al. (2003). The chemical network is taken from the
UMIST database (Millar et al. 1997; Le Teuff et al. 2000). We follow
the chemical evolution of 168 species involved in 1777 gas-phase and
grain reactions.  This paper presents the first results of such an extensive 
chemical model for shells in which OH is particularly abundant. Other 
potentially observable species were not considered by Elitzur \& de Jong (1978),
and Hartquist et al. (1995) did not give quantitative results, even for CH$_3$OH, relevant to the picture investigated here. The model calculation is carried out in two phases,
both of which are time-dependent. In Phase I, the clump of gas is formed
during the gravitationally driven collapse of an initially diffuse molecular
region from a hydrogen nucleus number density, n$_H$, to a much higher fixed
final number density. Free-fall collapse is assumed. During
this phase, gas-phase chemistry and freeze out on to dust grains with
subsequent processing (mainly hydrogenation) are assumed to occur (see
Viti et al. 2003 for a full description). The initial density of the
clump is taken to be 500 cm$^{-3}$ while the final density, n$_f$, is
treated as a free parameter. Note that the initial density is 
consistent with observations of the 
Rosette molecular cloud translucent clumps (Williams, Blitz \& Stark 1995). We have run a test model where the initial density was set to 100 cm$^{-3}$: the final abundances of the collapsed clump do not seem to be significantly affected. Hence, we have taken 500 cm$^{-3}$ as our initial density for all models.

The gas then remains motionless for some time and subsequently, in
Phase II, begins to move at a constant velocity, v, towards an
ionization front. The time for which the gas remains motionless is
determined by the extent to which we wish depletion to occur for the
particular run. At the time at which motion towards the front begins,
the visual extinction, A$_V$, of material between the location of the
parcel of gas under consideration and the ionization front is 10 mags.

   The relationship between the visual extinction and the distance to the
ionization front, z(t), at time t is taken to be
\begin{equation}
{\rm A}_V =  {\rm n}_H \times {\rm z(t)/ 1.6} \times {\rm 10} ^{21} {\rm cm} ^{-2}
\end{equation}

The radiation field at the ionization front is assumed to be $\chi$
times the standard unshielded mean interstellar radiation background
of Habing (1968). 

   At the time Phase I is over, a percentage, FR, of the nuclei of
elements more massive than helium initially in the gas phase has
frozen - out.  In Table~\ref{tb:frozen} we show the surface abundances of 
several
species at the beginning of Phase II for one of the models. During this latter phase, 
as the gas moves towards the ionization
front, A$_V$ decreases until the gas is unshielded enough (at
A$_{V,evap}$) that all surface material is returned instantaneously to
the gas phase. The assumption of instantaneous evaporation is motivated 
in this
context by the high radiation fields involved: the Viti et al. (2003) models
imply that the enhanced molecular
condensations ahead of Herbig-Haro objects, where the radiation field is higher than ambient
but certainly lower than  the radiation fields considered here, can only come from evaporated icy mantles which are rapidly injected into the gas phase.  

Thus the free parameters are n$_f$, FR, $\chi$, A$_{V,evap}$, and
v. A further free parameter is the fraction of CO converted into
methanol on the surface of the grains.  However, on the basis of
CH$_3$OH absorption observations (Menten et al. 1986) and OH maser models
(e. g. Gray et al. 1992) we know that CH$_3$OH has an abundance comparable to
or even greater than that of OH. There is no known gas phase mechanism at low
temperatures that produces quantities of CH$_3$OH comparable to those
found in CH$_3$OH maser sources. Surface chemistry is almost certainly
involved in the production of CH$_3$OH.

By running a model (Model 0, see Table 1) which simulates the maser
environment (n$_H$ = 10$^7$ cm$^{-3}$, T = 50 K; $\chi$ =
3$\times$10$^5$) we found that the CH$_3$OH column density is comparable to
the OH column density only if at least 25\% of the carbon monoxide sticking
on the grains is converted into methanol. Hence, for all the models presented
here we have assumed that 25\% of CO is converted into CH$_3$OH on grains.

The dynamical description that we have used is equally applicable
to a situation in which, in the frame of reference of the star, the
ionization front is expanding at speed v towards a gas that is
motionless, or to a situation in which the gravity of the star causes
gas to fall at a speed of v towards the star and approach an
ionization front that is not moving in the star's frame (cf. Keto
2002). In the Keto (2002) dynamical model of HII region, infall through a 
static front would maintain the overall dynamical and chemical structures 
near the front for close to the entire lifetime of an HII region.

We have explored a large parameter space resulting in 33
models. However, after a preliminary analysis, and for simplicity, we
decided to list here only the most plausible ones (see Table 1).
Note that we have retained only those models where Phase II begins with highly 
depleted material, which we believe is quite realistic because some dense cores, in 
regions where high mass stars have not started to form, have CO fractional abundances 
that are close to two orders of magnitude below those elsewhere (e.g. Caselli et al. 2002). 
Those cores almost certainly form on a timescale comparable to that of the free-fall time and may well remain quiescent for some time. The regions that we are modelling here have densities that are within a factor of a few to ten of those of the dense cores mentioned above and may well have dynamical and chemical histories similar to those of the dense cores as well.

\begin{table*}
\caption{Model Parameters. The notation a(b) signifies a $\times$ 10$^{b}$.
The model number is listed in Column 1; Column 2 shows the
density of the gas 
at the end of Phase I; Columns 3, 4 and 5 list respectively
the velocity of the clump, the strength of the radiation field and the visual
extinction at which the grains evaporate during Phase II. 
For all models listed here FR = 100\% and 25\% of CO is converted into CH$_3$OH on the grains.}
\begin{tabular}{|c|cccc|}
\hline
Model & $n_f$ (cm$^{-3}$) &  v (km/s) & $\chi$ & A$_{V_{evap}}$  \\
\hline
0  & 1(7) & 1 & 3(5) & 8 \\
1  & 1(5) &  1 & 3(3) & 8  \\
2  & 1(5) &  2 & 3(3) & 8  \\
3  & 1(5) &  1 & 3(4) & 8  \\
4  & 1(5) &  2 & 3(4) & 8  \\
5  & 1(5) &  1 & 3(3) & 6  \\
6  & 1(5) &  1 & 3(4) & 6  \\
7  & 1(4) &  1 & 3(3) & 6  \\
8  & 1(4) &  1 & 3(3) & 5  \\
9  & 1(5) &  1 & 3(4) & 5  \\

\hline
\end{tabular}
\end{table*}

\begin{table}
\caption{Surface fractional abundances of several
species at the beginning of Phase II for Model 8.}
\begin{tabular}{|c|c|}
\hline
H$_2$O & 4(-4) \\
CO & 1(-5) \\
H$_2$CO & 6(-8) \\
CH$_3$OH & 4(-6) \\
H$_2$S & 1(-5) \\
NH$_3$ & 3(-5) \\
CH$_4$ & 2(-4) \\
N$_2$ & 1(-5) \\
\hline
\end{tabular}
\label{tb:frozen}
\end{table}

\section{Results and Discussion}

Bourke et al. (2001) derived column densities for OH in the range of
4$\times$10$^{14}$ to 3$\times$10$^{15}$ cm$^{-2}$. For many models
for which FR is 25\% or 50\% (not shown), the OH column density
between A$_V$ = 10 and A$_V$ = A$_{V,evap}$ is greater than the OH column density for
A$_V$ less than A$_{V,evap}$. We wish to examine the possibility that
the absorption occurs primarily near the ionization front. Thus, we
consider here only models for which FR = 100\%.

Table 1 gives values of the parameters specifying each model. Table 2
gives the column density of various species between A$_V$ = 10 mags
(the beginning of Phase II) and A$_V$ equal to the value given in the
final column; in Figure 1 the fractional abundances of OH for selected
models are shown as a function of A$_V$, always after the grain mantles
have evaporated, up to A$_V$ $\sim$ 3.5 mags, corresponding to a
post-evaporation timescale of the order of 3$\times$10$^4$--3$\times$10$^5$ years.
\begin{table*}
{\small
\caption{Column densities of selected species for Models 1-9, calculated from 10 mags to the A$_V$ shown in column 11.}
\begin{tabular}{|c|ccccccccc|c|}
\hline
Model & OH & CO & HCO$^+$ & H$_2$CO & H$_2$S & SO & SO$_2$ & CH$_3$OH & NH$_3$ & A$_V$ \\
\hline
1  & 1.451(17)&  4.781(17)&  3.212(12)&  4.113(15)&  2.066(16)&  2.933(15)&  1.807(16)&  2.773(15)&  5.518(16)&2 \\
1  & 1.451(17)&  4.483(17)&  3.210(12)&  4.113(15)&  2.066(16)&  2.933(15)&  1.807(16)&  2.773(15)&  5.518(16)   & 4 \\
1  & 1.448(17)&  5.448(16)&  3.338(11)&  2.903(15)&  2.064(16)&  1.921(15)&  1.697(16)&  2.765(15)&  5.515(16)   & 6 \\
1  & ---       &  ---       &  ---       &  ---       &  ---       &  ---       &  ---       &  ---       &  ---          & 8 \\
2  & 1.835(17)&  4.379(17)&  3.647(12)&  4.699(15)&  2.705(16)&  3.089(15)&  1.858(16)&  3.584(15)&  7.300(16)   & 2 \\
2  & 1.835(17)&  3.745(17)&  3.645(12)&  4.699(15)&  2.705(16)&  3.089(15)&  1.858(16)&  3.584(15)&  7.300(16)   & 4 \\
2  & 1.731(17)&  3.253(16)&  4.375(10)&  2.431(15)&  2.670(16)&  7.600(14)&  1.242(16)&  3.493(15)&  7.216(16)   & 6 \\
2  & ---       &  ---       &  ---       &  ---       &  ---       &  ---       &  ---       &  ---       &  ---          & 8 \\
3  & 6.900(16)&  4.354(17)&  2.983(12)&  1.949(15)&  5.967(15)&  2.426(15)&  9.724(15)&  7.965(14)&  1.512(16)   & 2 \\
3  & 6.900(16)&  4.349(17)&  2.983(12)&  1.949(15)&  5.967(15)&  2.426(15)&  9.724(15)&  7.965(14)&  1.512(16) & 4 \\
3  & 6.899(16)&  2.121(17)&  2.964(12)&  1.940(15)&  5.967(15)&  2.426(15)&  9.724(15)&  7.965(14)&  1.512(16) & 6 \\
3  & ---       &  ---       &  ---       &  ---       &  ---       &  ---       &  ---       &  ---       &  ---       & 8 \\
4  & 1.094(17)&  4.008(17)&  3.603(12)&  2.525(15)&  9.391(15)&  2.847(15)&  1.208(16)&  1.245(15)&  2.420(16) & 2 \\
4  & 1.094(17)&  4.002(17)&  3.602(12)&  2.525(15)&  9.391(15)&  2.847(15)&  1.208(16)&  1.245(15)&  2.420(16) & 4 \\
4  & 1.094(17)&  1.419(17)&  3.292(12)&  2.415(15)&  9.391(15)&  2.846(15)&  1.208(16)&  1.245(15)&  2.420(16) & 6 \\
4  & ---       &  ---       &  ---       &  ---   &    ---       &  ---       &  ---       &  ---       &  ---        & 8 \\
5  & 3.377(16)&  3.625(17)&  2.066(12)&  1.720(15)&  1.766(15)&  1.300(15)&  4.637(15)&  2.837(14)&  4.934(15) & 2 \\
5  & 3.377(16)&  3.340(17)&  2.065(12)&  1.720(15)&  1.766(15)&  1.299(15)&  4.637(15)&  2.837(14)&  4.934(15) & 4 \\
5  & ---      &  ---      &  ---      &  ---      & ---       &  ---       &  ---       &  ---       &  ---        & 6 \\
5  & ---       &  ---       &  ---       &  ---   &    ---       &  ---       &  ---       &  ---       &  ---        & 8 \\
6  & 3.501(15)&  1.764(17)&  6.503(11)&  2.673(14)&  3.346(14)&  2.172(14)&  3.233(14)&  4.463(13)&  9.157(14) & 2\\
6  & 3.501(15)&  1.759(17)&  6.497(11)&  2.673(14)&  3.346(14)&  2.172(14)&  3.233(14)&  4.463(13)&  9.157(14) & 4\\
6  & --       & --       & --       & --       & --       & --       & --       & --       & --       & 6\\
6  & --       & --       & --       & --       & --       & --       & --       & --       & --       & 8\\
7  & 5.857(15)&  1.838(17)&  9.236(11)&  3.081(14)&  2.155(14)&  2.731(14)&  3.512(14)&  8.331(13)&  4.723(14) & 2 \\
7  & 5.857(15)&  1.833(17)&  9.230(11)&  3.081(14)&  2.155(14)&  2.731(14)&  3.512(14)&  8.331(13)&  4.723(14) & 4 \\
7  & 1.253(12)&  --      &  --       &  --       &  --       & --       & --       & --       &  --       &  6 \\
7  & 1.037(12)&  --       &  --       &  --       &  --       & --       & --       & --       &  --       &  8 \\
8   & 1.533(14)&  4.130(16)&  7.094(10)&  2.839(13)&  1.582(14)&  6.318(12)&  2.127(12)&  4.739(13)&  3.316(14)&  2 \\
8   & 1.532(14) & 4.086(16)&  7.035(10)&  2.838(13)&  1.582(14)&  6.303(12)&  2.127(12)&  4.739(13)&  3.316(14)&  4 \\
8   & 1.253(12)&  --       &  --       &  --       &  --       &  --       &  --       &  --       &  --       &  6 \\
8   & 1.037(12)&  --       &  --       &  --       &  --       &  --       &  --       &  --       &  --       &  8 \\
9  & 9.037(11)&  3.717(16)&  9.703(09)&  9.748(12)&  3.433(14)&  7.221(11)&  1.018(12)&  4.147(13)&  9.391(14) & 2 \\
9  & 8.849(11)&  3.672(16)&  9.118(09)&  9.737(12)&  3.433(14)&  7.066(11)&  1.018(12)&  4.147(13)&  9.391(14) & 4 \\
9  & ---       &  ---       &  ---       &  ---       &  ---       &  ---       &  ---       &  ---       &  ---        & 6 \\
9  & ---       &  ---       &  ---       &  ---       &  ---       &  ---       &  ---       &  ---       &  ---        & 8 \\
\hline
\end{tabular}
}
\end{table*}
 \begin{figure*}
   \centering
\includegraphics[angle=-90,width=18cm]{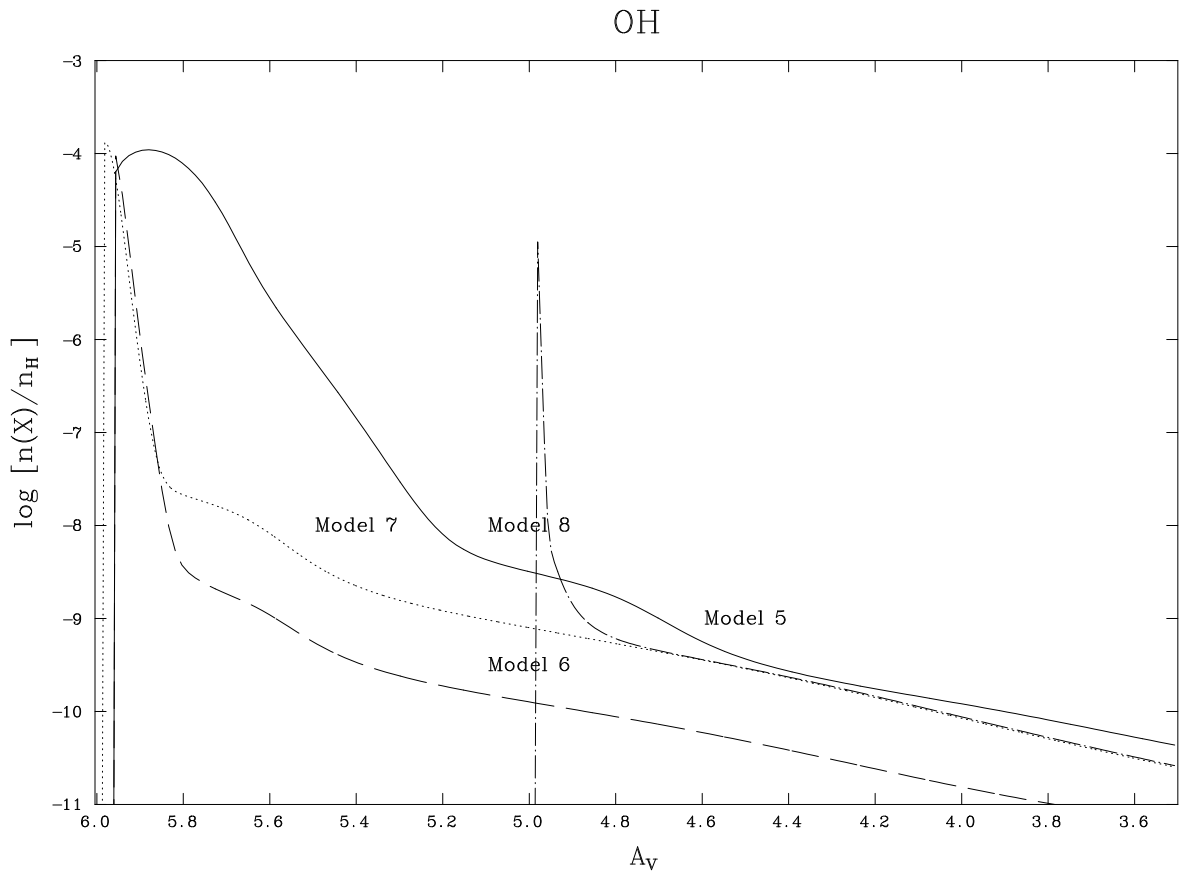}

   \caption{Fractional abundance of OH for selected models as a function of A$_V$.}
    \end{figure*}

From Figure 1, we see that soon after evaporation, OH is very abundant
for all models: this is due to the evaporation of water which
efficiently dissociates. OH is then destroyed, also by
photodissociation. We note that a high radiation field (e.g. Model 6) or a
low A$_{V,evap}$ (e.g. Model 8) accelerates the destruction of OH. Moreover,
another consequence of a high $\chi$ is also a very fast destruction of most
gas phase species.

From Table 2, we note that Models 6, 7 and 8, are the only ones for
which the OH column densities from A$_V$ = 0 to 10 mags are comparable
to those measured by Bourke et al. (2001). For all three of these
models v = 1 km s$^{-1}$. The values of $\chi$ are 3$\times$10$^4$ and
3$\times$10$^3$, which are within the range expected for early B and
late O type stars at the distances that the ionization fronts of
HII regions are from the central stars. The sensitivity
of the results to the adoption of different values of A$_{V,evap}$ is
demonstrated by a comparison of the results for Models 7 and
8. A$_{V,evap}$ has values of 6 and 5, respectively, for those
models. A value of A$_{V,evap}$ similar to these is compatible with
the inferences drawn by Viti et al. (2003) in their modelling of
chemistry triggered by UV radiation emitted in shocks associated with
Herbig - Haro objects.

\begin{figure*}
  \centering
\includegraphics[angle=0,width=18cm]{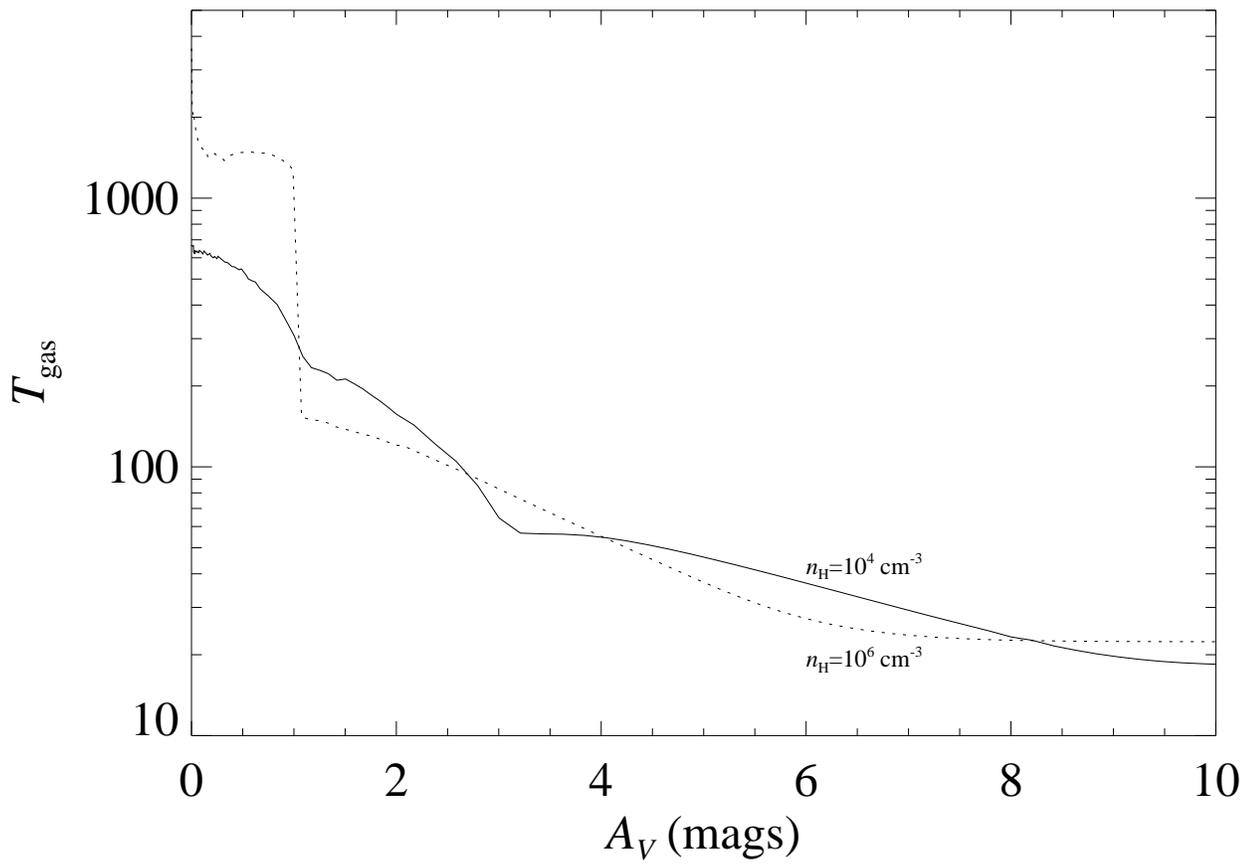}
   \caption{Gas temperature as a function of visual extinction for n$_H$ = 10$^4$ cm$^{-3}$, $\chi$ = 10$^3$ Habing (solid line) and n$_H$ = 10$^6$ cm $^{-3}$, $\chi$ = 10$^3$ Habing  (dotted line)}
   \label{fg:pdr}
    \end{figure*}

Little contribution to the OH column density comes from the region
where A$_V$ is less than 4 mags. Inspection of Figure 1 of Hartquist \&
Sternberg (1991) shows that the OH in the present models is, thus, in
regions where the temperature does not greatly exceed 10 K: this 
inference can be drawn solely from the high density results from Hartquist \&
Sternberg (1991) because for a fixed value of $\chi$/n, the temperature, as a function of A$_V$, does not vary significantly at A$_V$ $>$ 2 mags. For
fixed A$_V$ $<$ 1 mag and for fixed $\chi$/n, the temperature increases with density due mostly to the collisional de-excitation of radiatively pumped H$_2$ becoming more important with increasing density. Nevertheless, in order to 
confirm this inference, we ran two test models with the UCL PDR code 
(Bell et al. 2005; Papadopoulos et al. 2002) at densities of 10$^4$ and 10$^6$ cm$^{-3}$.
The results of these tests are shown in Figure~\ref{fg:pdr}. From this figure, 
it is clear that, while the temperature is high at the outer 
edge, at A$_V$ $>$ 4 mags, the temperature 
has values well below 100K and close enough to 10 K for this assumption to be appropriate for the chemical calculations.
Note that the UCL PDR code and the chemical code
used for the calculation of the grids reported in this paper are two versions of the same
basic code. The main difference is, of course, in the treatment of the temperature: 
it is given
as an input in the chemical code, while
it is self-consistently computed in the UCL PDR code. The latter code has been recently benchmarked against many other PDR codes 
(Roellig et al. in prep).  

Clearly the highest fractional abundances of OH and of other species
is obtained only in a narrow A$_V$ range near A$_{V,evap}$. This
result justifies our adoption of plane parallel geometry.

If the observed OH towards the Bourke et al. sources is indeed caused
by grain evaporation in parcels of gas moving toward the ionization
front, then it would be desirable to find other tracers of this
process. In particular, we note that CH$_3$OH should also be enhanced
as a direct consequence of grain evaporation. From Table 2, it is clear
that CH$_3$OH is always less abundant than OH for models 6 to 8, though
in Model 0, which gives a column density of OH in the range of those
in OH masers, the CH$_3$OH and OH column densities are similar. For models
6 to 8, the highest column density of CH$_3$OH is reached for Model 7,
while for these three models the densities of OH and CH$_3$OH differ by the
smallest factor in Model 8.

\section{Conclusions}

In this paper we have put forward and explored the possibility that
the OH absorption observed toward HII regions is due to the
evaporation of grain mantles primarily near the HII regions' ionization
fronts.

We show that it is possible for most of the detected OH to be in such
locations provided that the grain mantles evaporate $before$ the gas
becomes completely unshielded to the strong radiation fields typical
of these environments. In order to test this idea, additional absorption 
observations are desirable for the lines-of-sight for which 
Zeeman measurements of magnetic fields
are made. They should be
made in the lines of species that are often not abundant in cold dark
cloud material but are abundant in the models presented here that give
OH column densities near those measured by Bourke et al. (2001). In
this respect CH$_3$OH and H$_2$S are particularly promising species,
although a high column density of methanol is only obtained if a
substantial amount of CO is converted into CH$_3$OH on the grains
before evaporation.   
Methanol is routinely observed towards regions of high mass star
formation, particularly towards UCHII regions, where it is mainly detected
in maser emission (e. g. Walsh et al. 1997), and hot cores.
H$_2$CO is also regularly observed in absorption, 
and the model results suggest that it too would be a good candidate
for observations in appropriate directions.  
In fact, Downes et al. (1980) reported low spatial resolution observations
of H$_2$CO towards several galactic sources and found that, at least in some
cases, it is seen at the same velocity as the OH absorption.
The high abundance of CH$_3$OH, H$_2$S and H$_2$CO are all a consequence of hydrogenation of simpler species on the grains during the collapse phase; once A$_{V, evap}$ is reached,
hydrogenated species in the gas phase are enhanced. In the case of methanol, there is an additional contribution from the high radiation field which causes
a high abundance of CH$_3^+$; the latter efficiently reacts with water (also enhanced on the grains during the collapse phase) and forms the ion CH$_3$OH$_2^+$ 
which then produces methanol via electronic recombination.
\section{Acknowledgements}
SV acknowledges individual financial support from a PPARC Advanced Fellowship.
The collaboration was supported by a PPARC Visitors Grant held in Leeds.
\section{References}

Allen, A., Shu, F. H. 2000, ApJ, 536, 368 \\
Bell, T, Viti, S, Williams, D A, Crawford, I A, Price R J, 2005, 357, 961 \\
Bourke, T. L., Myers, P. C., Robinson, G., Hyland, R. 2001, ApJ, 554, 916 \\
Caselli, P., Walmsley, C. M., Zucconi, A., Tafalla, M., Dore, L., Myers, P. C., 2002, ApJ, 565, 344 \\
Crutcher, R. M. 1999, ApJ, 520, 706 \\
Downes, A. J. B., Wilson, T. L., Bieging J., Wink, J., 1980, A\&AS, 40, 379 \\
Elitzur, M., de Jong, T. 1978, A\&A, 67, 323 \\
Gray, M. D, Field, D., Doel, R. C. 1992, A\&A, 262, 555 \\
Habing, H. J. 1968, BAN, 20, 177 \\
Hartquist, T. W., Sternberg, A. 1991, MNRAS, 248, 48 \\
Hartquist, T. W., Menten, K. M., Lepp, S., Dalgarno, A. 1995, MNRAS, 272,
    184 \\
Keto, E. 2002, ApJ, 580, 980  \\
Le Teuff, Y. H., Millar, T. J., Markwick, A. J. 2000, A\&AS, 146, 157 \\
Menten, K. M., Walmsley, C. M., Henkel, C, Wilson, T. L. 1986, A \&A, 157, 318 \\
Menten, K. M., Reid, M. J., Pratap, P., Moran, J. M.,  Wilson, T. L. 1992, 
    ApJ, 401, L39 \\
Millar, T. J., Farquhar, P. R. A.,  Willacy, K. 1997, A\&AS, 121, 139 \\
Mouschovias, T. Ch., Spitzer, L. Jr. 1976, ApJ, 210, 236 \\
Papadopoulos, P, P, Thi, W.-F.; Viti, S, 2002, ApJ, 579, 270 \\
Shu, F. H., Adams, F. C., Lizano, S. 1987, ARA\&A, 25, 23\\
Shu, F. H., Allen, A., Shang, H., Ostriker, E. C.,  Li, Z.-Y. 1999, in The 
   Origin of Stars and Planetary Systems, ed. C. J. Lada \& N. D. Kylafis 
   (Dordrecht: Kluwer), 193 \\
Stantcheva, T., Caselli, P., Herbst, E. 2001, A\&A, 375, 673 \\
Viti, S., Girart, J. M., Garrod, R., Williams, D. A., Estalella, R. 2003, 
   A\&A, 399, 187 \\ 
Walsh, A. J., Hyland A. R., Robinson G., Burton M. G., 1997, MNRAS, 291, 261 \\
Williams, J. P., Blitz, L., Stark, A. A., 1995, ApJ, 451, 252 \\
\end{document}